\documentclass{nature}

\usepackage{graphicx}

\title{A ground-based transmission spectrum of the super-Earth exoplanet GJ\,1214b}

\author{Jacob L.~Bean$^{1,2}$, Eliza Miller-Ricci Kempton$^{3}$, \& Derek Homeier$^{2}$}

\begin{document}

\maketitle

\begin{affiliations}
 \item Harvard-Smithsonian Center for Astrophysics, 60 Garden Street, Cambridge, MA 02138, USA
 \item Institut f\"ur Astrophysik, Georg-August-Universit\"at, Friedrich-Hund-Platz 1, 37077 G\"ottingen, Germany
 \item Department of Astronomy and Astrophysics, University of California, Santa Cruz, CA 95064, USA
\end{affiliations}

\begin{abstract}
In contrast to planets with masses similar to that of Jupiter and higher, the bulk compositions of planets in the so-called super-Earth regime ($M_{p}$ = 2 -- 10\,M$_{\oplus}$) cannot be uniquely determined from a mass and radius measurement alone. For these planets, there is a degeneracy between the mass and composition of the interior and a possible atmosphere in theoretical models\cite{adams08,rogers10a}. The recently discovered transiting super-Earth GJ\,1214b is one example of this problem\cite{charbonneau09}. Three distinct models for the planet that are consistent with its mass and radius have been suggested\cite{rogers10b}, and breaking the degeneracy between these models requires obtaining constraints on the planet's atmospheric composition\cite{millerricci09,millerricci10}. Here we report a ground-based measurement of the transmission spectrum of GJ\,1214b between 780 and 1000\,nm. The lack of features in this spectrum rules out cloud-free atmospheres composed primarily of hydrogen at 4.9\,$\sigma$ confidence. If the planet's atmosphere is hydrogen-dominated, then it must contain clouds or hazes that are optically thick at the observed wavelengths at pressures less than 200\,mbar. Alternatively, the featureless transmission spectrum is also consistent with the presence of a dense water vapor atmosphere. 
\end{abstract}

We observed transits of the planet GJ\,1214b in front of its host star on UT dates 29 April and 6 June 2010 using the FORS2 instrument on the UT1 telescope of the Very Large Telescope facility. The instrument was configured for multi-object spectroscopy using a mask with slits positioned on GJ\,1214 and six other nearby reference stars of similar brightness. The slits had lengths of 30 arc seconds and widths of 12.0 arc seconds to eliminate possible differential losses due to variations in the telescope guiding and seeing\cite{knutson07}. Complete spectra from 780 -- 1000\,nm with a resolution of approximately 1\,nm were obtained for all the stars in each exposure. A total of 197 exposures were obtained during the two observing runs, 88 of which were during a transit.

We extracted both ``white'' and eleven 20\,nm channel width spectrophotometric light curves for GJ\,1214 and the reference stars by summing the obtained spectra over wavelength. We corrected the transit light curves by combining the fluxes for five of the reference stars and dividing them from the flux of GJ\,1214. After this reduction, the light curves for GJ\,1214 exhibit the expected transit morphology superimposed on a curvature that is well-matched by a second order polynomial as a function of time. We modeled this trend for each time series simultaneously with the transit modeling described below. Normalized and corrected light curves for the spectrophotometric data are shown in Figure 1. The photon-limited uncertainties in the measurements after reduction and correction are 350 -- 710\,ppm. We found that these estimates potentially underestimate the true uncertainties in the data, and we therefore revised the uncertainties upward by 25 -- 78\% to yield reduced $\chi^{2}$ values of unity for the light curve model fits. See the Supplementary Information for more details on the observations, data reduction, and reference star corrections.

We fitted models for the transit\cite{mandel02} to the spectrophotometric time-series data to measure the apparent radius of GJ\,1214b in each channel - this is the planet's transmission spectrum. We assumed the planet is on a circular orbit, and we fixed the transit and system parameters to the values we determined from an analysis of the white light curves. The uncertainties in the fixed parameters do not contribute any additional uncertainty in the transmission spectrum, and only the overall level of the transmission spectrum would be influenced by changing these values within their confidence limits. 

We detect a marginally significant (1.5\,$\sigma$) difference in the depths of the white light curves between the two observed transits of 2.4 $\pm$ 1.6 x 10$^{-4}$. This level of variation is consistent with brightness variations of the unocculted portion of the stellar disk similar to the contemporaneously observed I-band variation of 0.65\% (Z. Berta personal communication). We accounted for this variability in our analysis of the spectrophotometric light curves by reducing the planet-to-star radius ratio parameter used to generate the models for the first transit by an amount corresponding to the change in depth in the white light curves ($\Delta\,R_{p}/R_{\star}$\,=\,0.0010). We did not apply any additional corrections to account for a color bias due to occulted or unocculted spots because simulations we performed indicate that the effects across the small wavelength range studied here would be negligible given the precision of the data. 

The results from our analysis of the white light curves after the stellar variability correction confirm and refine the previous estimates of system parameters. Assuming the mass of the star is 0.157 $\pm$ 0.019\,M$_{\odot}$\cite{charbonneau09}, the radius of the star is 0.206 $\pm$ 0.009\,$R_{\odot}$ and the average radius of the planet is 2.63 $\pm$ 0.11\,R$_{\oplus}$. The planet's average central transit time is BJD 2455315.794502 $\pm$ 0.000047 (4\,sec uncertainty), and its orbital period is 1.58040834 $\pm$ 0.00000034\,d. The determined transmission spectrum for GJ\,1214b in terms of the planet's radius assuming the stellar radius from the analysis of the white light curves is shown in Figure 2. We obtain similar results for the transmission spectrum when repeating the analysis with different plausible values for the fixed parameters, not adopting a prior on the limb darkening, applying no activity corrections, applying higher-order activity corrections, or using different approaches to correct for the activity. We also obtain the same results when analyzing the data for each transit separately. 

Previous work has shown that GJ\,1214b must have a significant atmosphere because its density is too low for it to be composed only of solid material\cite{rogers10b}. However, the composition of the atmosphere can not be inferred with only the knowledge of the planet's mass and radius due to degeneracies between interior structure and atmospheric models. Comparison of our observed transmission spectrum to model spectra for different atmospheric compositions (see Figure 2) indicates that we would have detected significant variations in the planet's effective radius, due to absorption of starlight by water vapor in the limb of the planet, if the atmosphere was mainly composed of hydrogen because such an atmosphere would have a large scale height. A hydrogen-dominated atmosphere would be expected if the planet was a scaled down version of the Solar System ice giants Uranus and Neptune with a primordial atmosphere accreted from the protoplanetary nebulae, or if it was a rocky planet that had outgassed large quantities of hydrogen during formation or a period of tectonic activity\cite{elkinstanton08}. Instead, we detect no such variations at high confidence (4.9\,$\sigma$), and we conclude that the planet must possess something other than a simple cloud-free hydrogen-dominated atmosphere. 

Of the models proposed for the planet based on constraints from interior modeling\cite{rogers10b,millerricci10,nettelmann10}, only the predicted spectra from cloud-free atmospheres composed predominantly of water vapor (steam) agree with our measured transmission spectrum. Water vapor is present in the planet's limb to absorb the starlight in all the proposed models, but at least 70\% water vapor by mass (mean molecular weight $>$\,5$M_H$, where $M_H$ is the mass of the hydrogen atom) is needed to result in a transmission spectrum with the relatively small variations required to be consistent with our measurements within 1\,$\sigma$. A predominantly steam atmosphere is a component of the ``water world''\cite{kuchner03,leger04,selsis07} bulk composition model of the planet discussed in Ref.~4 (their case II), although it has been recently suggested\cite{nettelmann10} that a water-rich atmosphere is not necessarily indicative of a water-rich interior. In either case, the planet would not harbor any liquid water due to the high temperatures present throughout its atmosphere\cite{rogers10b,nettelmann10}. If GJ\,1214b contains a significant water abundance, then it likely formed beyond the snow line of its host star's protoplanetary disk, and subsequent orbital evolution brought it inward to its current location. The planet either could have not accreted as much nebular gas (which would have been composed primarily of hydrogen) as Neptune in the first place, or it could have subsequently lost hydrogen-rich gas that it did accrete due to atmospheric escape\cite{charbonneau09,rogers10b}. 

Another possible explanation for the lack of spectral features observed in the transmission spectrum of GJ\,1214b is that the planet could have a high layer of clouds or hazes, obscuring the view of lower regions in the planetary atmosphere. The effect of such a cloud deck on the transmission spectrum would be to reduce the strength of predicted spectral features, even for an atmosphere with a low mean molecular weight. We performed a simple test to determine the effect of clouds on the transmission spectrum for the case of a hydrogen-dominated atmosphere by cutting off all transmission of starlight below the height of a hypothetical cloud deck. We find that clouds or hazes located at pressures less than 200\,mbar (comparable to cloud/haze layers on Venus and Titan) do flatten out the transmission spectrum of a hydrogen-dominated atmosphere to the point where it is indistinguishable from a steam atmosphere using the current data. Therefore, our observations could be indicative of clouds or hazes in the upper parts of GJ\,1214b's atmosphere. Although no candidate has yet been presented for a cloud layer on GJ\,1214b, and the planet's proposed temperature-pressure profile does not cross the condensation curves of any known equilibrium condensates that are likely to be present in significant quantities \cite{lodders06}, hazes resulting from photochemical processes remain a viable prospect and are currently unconstrained by models. Transit observations of GJ\,1214b in the infrared could serve to shed light on the question of clouds and hazes in the planet's atmosphere, since scattering from cloud and haze particles is far less efficient at these wavelengths, and radius variations in the planet's transmission spectrum could be more apparent if the atmosphere was actually composed primarily of hydrogen. Indeed, a recent study has suggested there are variations in the planet's transmission spectrum between the near-infrared J and K bands that are qualitatively consistent with the expectations for a hydrogen-dominated atmosphere (B. Croll personal communication). Further observations are needed to clarify this issue.

The only other known transiting planet in the same mass regime as GJ\,1214b, CoRoT-7b\cite{leger09,queloz09}, could also harbor an atmosphere despite the extreme level of insolation it receives from its host star\cite{valencia10}. However, this planet orbits a much larger star ($R_{\star}$ = 0.87\,$R_{\odot}$, Ref.~14) than GJ\,1214b and, thus, it is unlikely that additional constraints on its atmosphere can be obtained with existing facilities due to the unfavorable planet-to-star radius ratio. In contrast, our results confirm previous predictions\cite{deming09,kaltenegger09} about the excellent prospects for the detailed characterization of transiting exoplanets identified by searches targeting very low-mass stars like the MEarth project\cite{nutzman08} and near-infrared radial velocity surveys\cite{bean10}. The increasingly sophisticated observational techniques for transiting exoplanet spectroscopy as applied to study the atmospheres of planets orbiting M dwarfs offers the promise of comparative studies of super-Earth type planets in the near-future and, ultimately, probably the best chance for the eventual first characterization of the atmosphere of a potentially habitable planet.

\bibliography{ms.bib}

\begin{addendum}
\item [Supplementary Information] is linked to the online version of the paper at www.nature.com/nature.
\item We thank David Charbonneau, Jean-Michel Desert, Jonathan Fortney, Sara Seager, Leslie Rogers, and Dimitar Sasselov for discussions about this work. J.L.B. received funding from the European Commission’s Seventh Framework Program as a Marie Curie International Incoming Fellow. J.L.B. and E.M.-R.K. acknowledge funding from NASA through the Sagan Fellowship Program. The results presented are based on observations made with ESO Telescopes at the Paranal Observatories under program IDs 284.C-5042 and 285.C-5019.
\item[Author Contributions] J.L.B. performed the observations and data analysis, and led the overall direction of the project. E.M.-R.K. calculated theoretical models for the planetary atmosphere. D.H. calculated the stellar limb darkening. J.L.B. and E.M.-R.K. wrote the telescope time proposals and the paper. All authors discussed the results and commented on the manuscript.
\item[Author Information] The data utilized in this work can be accessed at the ESO/ST-ECF science archive (archive.eso.org/cms/). Reprints and permissions information is available at\\
npg.nature.com/reprintsandpermissions. The authors declare that they have no competing financial interests. Correspondence and request for materials should be addressed to J.L.B. (jbean@cfa.harvard.edu).
\end{addendum}

\begin{figure*}
\resizebox{14.0cm}{!}{\includegraphics{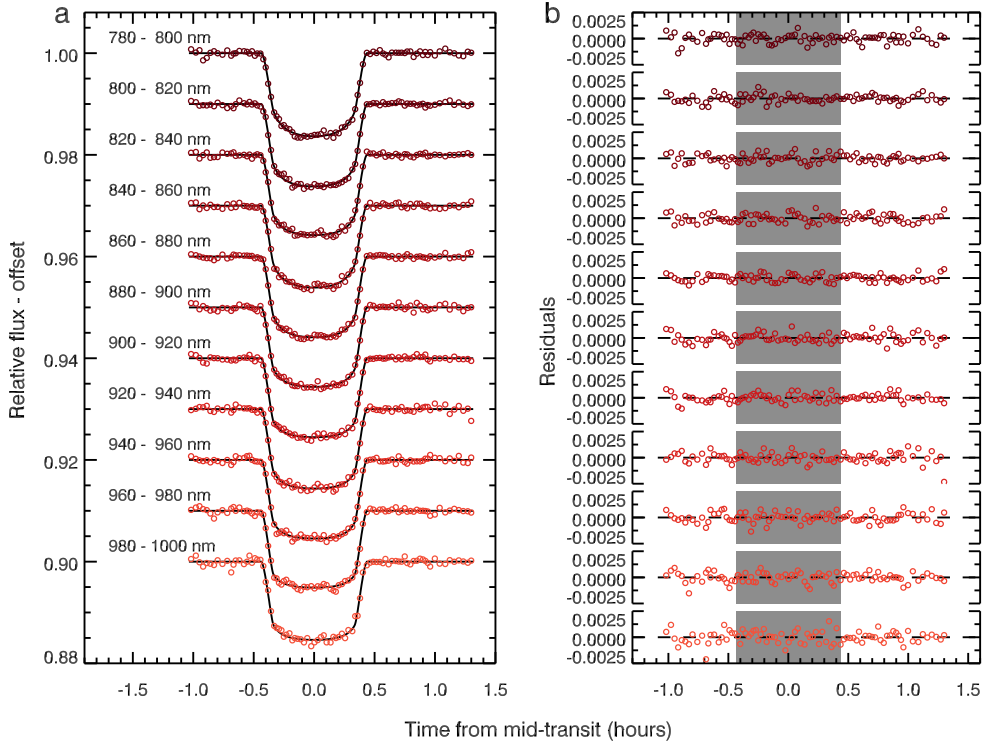}}
\caption{\textbf{Spectrophotometric data for transits of GJ\,1214b} The free parameters in the model fitted to the data to determine the transmission spectrum were the ratio of the radius of the planet to the radius of the host star, quadratic limb darkening coefficients, and the three terms describing the second order polynomial as a function of time needed to normalize the data. The ratio of the radii and the limb darkening coefficients were determined for each channel, while the polynomial coefficients for the data normalization were determined for each channel in each transit observation. We estimated limb darkening coefficients that were used as priors with uncertainties for each spectrophotometric channel using a spherically symmetric model atmosphere for the host star calculated with the PHOENIX code\cite{hauschildt99} for parameters $T_{eff}$ = 3026\,K, [M/H] = 0.0, log\,g = 5.0, and $R_{\star}$ = 0.21\,$R_{\odot}$. The model intensities were integrated over the bandpass for each channel to account for the variation of the instrument transmission function and the spectral energy distribution of the star. To estimate the uncertainty in these predictions, we repeated the calculations for models with $T_{eff}$ = 3156 and 2896\,K. The normalized light curves and best-fit models are shown in \textbf{a}. An offset of 0.01 was subtracted from the data in each successive channel for clarity. The data from the two observed transits were combined after correcting for the variation in the transit depth between the observed transits (the correction factor used was 1.017), and binned at a sampling of 72\,s for this figure. The solid lines show the best-fit models fitted independently to the data to determine the apparent size of the planet in each channel. The binned residuals from the best-fit models are shown in \textbf{b}. The grey boxes indicate the times when the planet is passing in front of the star. The standard deviation of the residuals in the channels ranges from 331 to 580\,ppm.}
\end{figure*}

\begin{figure*}
\resizebox{16.0cm}{!}{\includegraphics{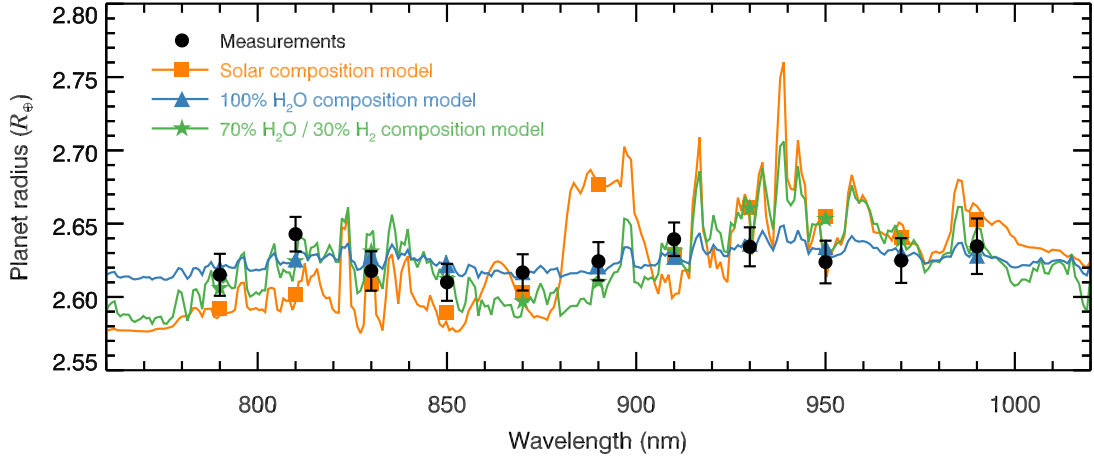}}
\caption{\textbf{The transmission spectrum of GJ\,1214b compared to models} Theoretical predictions of the transmission spectrum for GJ\,1214b\cite{millerricci10} are shown for atmospheres with a solar composition (i.e., hydrogen-dominated; orange line and squares), a 100\% water vapor composition (blue line and triangles), and a mixed composition of 70\% water vapor and 30\% molecular hydrogen by mass (green line and stars). The points for the models give the expected values for the transmission spectrum in each of the spectrophotometric channels. All of the features in the model spectra arise from variations in the water vapor opacity, with the exception of the feature at 890\,nm that is due to methane absorption. The measurements and their uncertainties (black circles) were estimated by fitting the spectrophotometric data using five Markov chains with 2.5 x 10$^{5}$ steps. The uncertainties, which are valid for the relative values only, are the 1\,$\sigma$ confidence intervals of the resulting posterior distributions, and are consistent with the estimates we obtained from a residual permutation bootstrap analysis. The uncertainty in the absolute level is 0.11\,$R_{\oplus}$, and is due mainly to the uncertainty in the host star mass. The data are consistent with the model for the water vapor atmosphere ($\chi^{2}$\,=\,5.6 for 10 degrees of freedom) and inconsistent with the model for the solar composition model at 4.9\,$\sigma$ confidence ($\chi^{2}$\,=\,47.3). The predictions for a solar composition atmosphere with CH$_{4}$ removed due to photodissociation (not shown) are equally discrepant with the data. The mixed water vapor and molecular hydrogen atmosphere model contains the most hydrogen possible to still be within 1\,$\sigma$ ($\chi^{2}$\,=\,11.5) of the measurements. The data are furthermore consistent with a hydrogen-dominated atmosphere with optically thick clouds or hazes located above a height of
200\,mbar (not shown). We obtain similar results when comparing the models to measurements obtained using smaller or larger channel sizes.}
\end{figure*}

\newpage

\begin{center}
{\LARGE{\bf Supplementary Information}}
\end{center}

\noindent In this supplement to the article reporting a ground-based transmission spectrum of the super-Earth planet GJ\,1214b, we describe in more detail the observational, data reduction, and photometric correction methods.

\vspace{-4mm}

\section*{Observations}\vspace{-4mm}
We obtained time-series observations during two transits of the planet GJ\,1214b in front of its host star (the transit duration is 52 minutes) using the FORS2 instrument on the UT1 telescope of the Very Large Telescope facility at the European Southern Observatory on Cerro Paranal in Chile. The instrument was configured for medium-resolution multi-object spectroscopy using a mask with slits positioned on GJ\,1214 and six other nearby reference stars of similar brightness (the range in the ``white'' instrumental magnitudes is 2.7). The slits had lengths of 30.0 arc seconds and widths of 12.0 arc seconds. The instrument de-rotator was used to maintain a constant alignment of the slits on the stars throughout the observations. We employed the 600z grism for the dispersive element and we used the OG590 filter to isolate the first order spectra. The instrument was configured the same way for both transits and the same slit mask was used. Complete spectra from 780 -- 1000\,nm for all the stars were imaged on to a mosaic of two CCD detectors during each exposure. Except for the first few exposures during the first transit, all exposure times were 35\,s, which gave a cadence of 72\,s including the read and reset time of the detector. A total of 197 exposures were obtained, 88 of which were during a transit.

The observations during the first transit (UT 29 April 2010) were performed for 119 minutes. The observations began 15 minutes before the transit ingress and continued for 52 minutes after the transit egress. The field was at an airmass of 1.21 at the beginning of the observations, rose to an airmass of 1.15, and then set to an airmass of 1.17 by the end. The conditions were clear, but the seeing was unstable and varied between 0.6 and 1.6 arc seconds as measured from the widths of the spatial profiles in the obtained spectra. Some of the pixels used to record the spectrum of the the brightest star that was observed (one of the reference stars) were saturated during the third and fourth exposures in the sequence. These exposures were not utilized in our analysis. 

The observations during the second transit (UT 6 June 2010) were also performed for 119 minutes. The observations began 36 minutes before the transit ingress and continued for 31 minutes after the transit egress. The field was at an airmass of 1.21 at the beginning of the observations, rose to an airmass of 1.15, and then set to an airmass of 1.18 by the end. The conditions were again clear, but with seeing that varied between 0.9 and 2.4 arc seconds. None of the exposures contained saturated pixels.

\vspace{-4mm}

\section*{Data reduction and spectral extraction}\vspace{-4mm}
After standard CCD reduction steps including bias subtraction and flat fielding, we extracted one-dimensional spectra from the recorded images and estimated their corresponding uncertainties using the optimal extraction algorithm (Horne 1986, PASP, 98, 609). The background and its variance were estimated and subtracted for each spectrum individually during the extraction based on the observed flux in the illuminated regions of the images that were separated by 7.5 -- 15.0 arc seconds from the stellar spectra in the spatial direction. 
Detector pixels that were either bad, or that were affected by cosmic ray strikes in individual images were identified and masked for the spectral extraction.

\begin{figure*}
\resizebox{16.0cm}{!}{\includegraphics{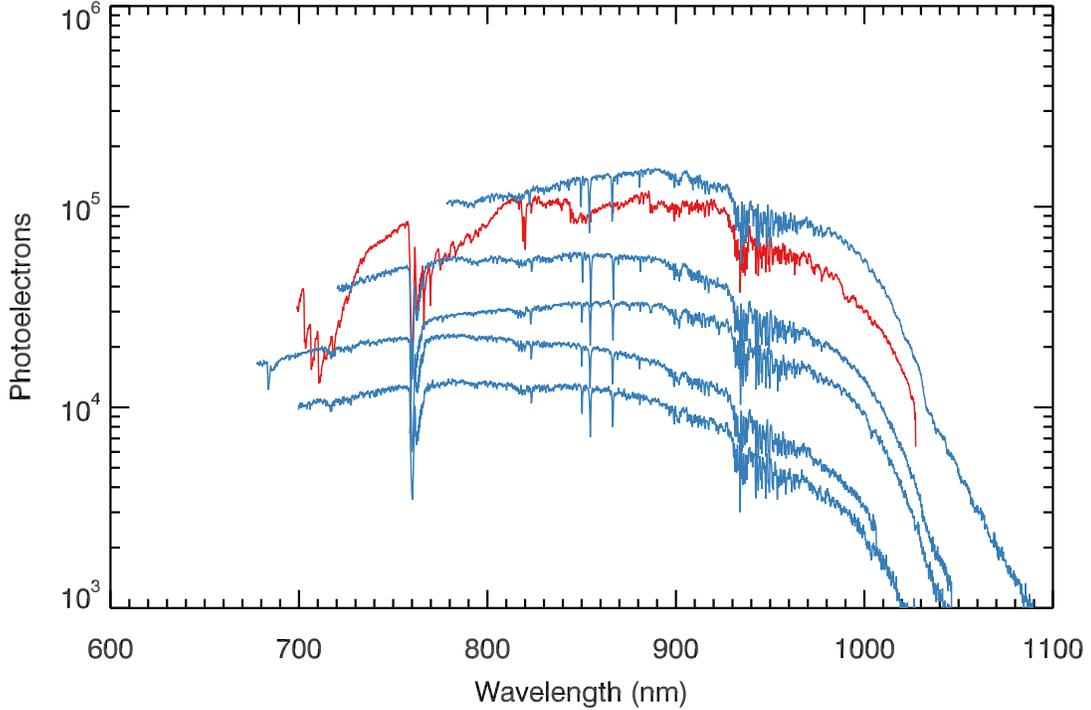}}
\caption{Example extracted spectra for GJ\,1214 (red) and the five reference stars utilized for the relative photometric calibration (blue). The different wavelength coverage for the objects is a result of parts of their spectra not falling on the detector due to their distribution on the sky in the dispersion direction.}
\end{figure*}

We established the wavelength calibration of the extracted stellar spectra based on spectra of an emission lamp obtained following each transit observation using a copy of the science mask with slit widths set to 1.0 arc second. We adopted a linear form for the wavelength solution. The typical value of the dispersion term was 0.160\,nm\,pixel$^{-1}$. The wavelength calibration was done independently for each of the objects, although only the zero-point of the wavelength scale varied significantly. We estimate that the useful resolution of the stellar spectra is approximately 1\,nm including the intrinsic resolving power of the grism (R $\equiv$ $\lambda$/$\Delta \lambda$ $\approx$ 1390 for a 1.0 arc second slit), the atmospheric seeing, the motion of the stars on the slits during the observations (less than 0.2 arcseconds for each sequence), and the uncertainty in the positioning of the calibration slit mask relative to the science mask in the instrument.

Example spectra for GJ\,1214 and the five reference stars utilized for the relative photometry correction are shown in Figure S1. The typical signa-to-noise ratio (SNR) of the GJ\,1214 spectra is 250\,pixel$^{-1}$. The typical SNR for the reference stars ranges from 70 to 290\,pixel$^{-1}$.

\vspace{-4mm}

\section*{Photometric corrections}\vspace{-4mm}
As described in the main text, we summed the obtained spectra for GJ\,1214 and the reference stars over wavelength to create photometric time-series. Both a ``white'' time-series based on the full wavelength coverage, and 20\,nm wide (125 pixels) spectrophotometric time-series were created. Two corrections were applied to these time series for GJ\,1214 before they were compared to transit light curve models. Both of the corrections utilized are similar to standard reductions applied to aperture photometry performed on images to generate relative photometry. Figure S2 illustrates the step-by-step corrections applied to one of the spectrophotometric channel time-series.

\begin{figure*}
\resizebox{16.0cm}{!}{\includegraphics{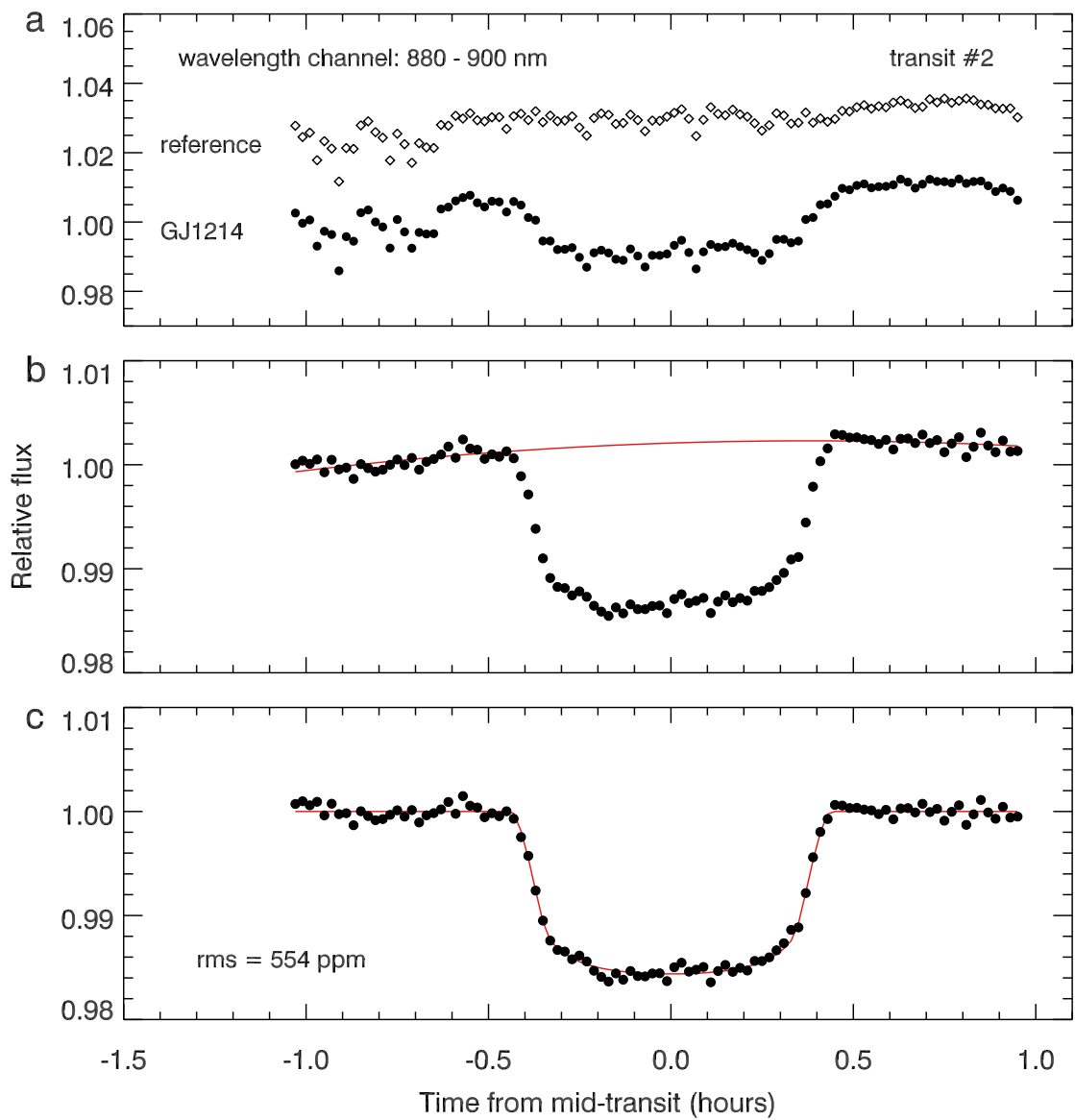}}
\caption{Illustration of the correction procedure for one of the spectrophotometric channels. Panel \textbf{a} shows the normalized light curves of GJ\,1214 (circles) and the composite of the five reference stars used to correct the data (diamonds). Panel \textbf{b} shows the light curve for GJ\,1214 after dividing out the reference star composite. The red line shows the fit for the slow curvature as a function of time determined simultaneously with the light curve modeling. Panel \textbf{c} shows the light curve for GJ\,1214 after dividing out the slow trend. The red line shows the determined transit model.}
\end{figure*}

The first correction we applied was to use the reference star time series to remove the effects of atmospheric transparency variations. To do this, we summed the fluxes of five of the reference stars and divided them from the flux of GJ\,1214. One of the observed reference stars (the faintest one) was not utilized because it yielded a slightly worse correction as evidenced by an increase in the out-of-transit residuals in the final light curve fits when it was included in this step. The uncertainties in the reference star fluxes were propagated through to the resulting flux for GJ\,1214. The reference star corrections were done channel-by-channel. That is, we summed the spectra over wavelength first rather applying a correction pixel-by-pixel (note that the spectra do have have the same sampling).

After the reference star correction, the time-series of GJ\,1214 exhibit the expected transit morphology superimposed on a slow curvature with time. This is likely an effect related to the color difference between GJ\,1214 and the reference stars. The effect is nearly linear with respect to airmass, but an additional term in either time or airmass significantly improves the fit quality. Ultimately, we modeled this effect as a second order polynomial with time simultaneously with the transit light curve modeling. Therefore, the uncertainty in the transmission spectrum accounts for the additional uncertainty from the covariance with this correction. The data exhibit no remaining correlation with airmass, seeing, or position of the spectrum in the two dimensional images after the reference star and slow trend corrections.

\end{document}